\documentclass[twocolumn,showpacs,amsmath,amssymb,nofootinbib]{revtex4-1}
\usepackage{dcolumn}
\usepackage{bm}
\usepackage{graphicx}

\newcommand{\be}{\begin{equation}}
\newcommand{\ee}{\end{equation}}
\newcommand{\ba}{\begin{eqnarray}}
\newcommand{\ea}{\end{eqnarray}}
\newcommand{\non}{\nonumber}
\newcommand{\n}[1]{\label{#1}}
\newcommand{\eq}[1]{(\ref{#1})}

\newcommand{\hhh}{\, ,\hspace{0.5cm}}
\newcommand{\BM}[1]{{\mbox{\boldmath $#1$}}}
\newcommand{\bi}[1]{\bibitem{#1}}

\newcommand{\hal}{\hat{\alpha}}
\newcommand{\hmu}{\hat{\mu}}
\newcommand{\hnu}{\hat{\nu}}
\newcommand{\hv}{\hat{v}}
\newcommand{\hgam}{\hat{\gamma}}
\newcommand{\htt}{\hat{t}}
\newcommand{\hT}{\hat{T}}
\newcommand{\hR}{\hat{R}}

\begin{document}

\title{Synchrotron Radiation From Weakly Magnetized Schwarzschild Black Hole}
\author{Andrey A. Shoom}
\email{ashoom@ualberta.ca}
\affiliation{Theoretical Physics Institute, University of Alberta,
Edmonton, Alberta, T6G 2E1, Canada\\ Department of Mathematics and Statistics, Memorial University, St. John's, Newfoundland and Labrador, A1C 5S7, Canada}

\begin{abstract}
We consider a synchrotron radiation from a charged particle moving in a bound orbit around a weakly magnetized Schwarzschild black hole (a static black hole immersed into a constant uniform magnetic field) in its equatorial plane, perpendicular to the magnetic field. In particular, we study the case when the Lorentz force acting on the charged particle is directed outward from the black hole. The particle is initially moving in a nongeodesic bound orbit which due to synchrotron radiation decays to a nongeodesic circular orbit. We study this transition and calculate the radiated power and energy loss of the particle.  

\end{abstract}


\maketitle

\section{Introduction}

Recent observations indicate that there is a strong magnetic field of several hundred gauss around a supermassive  black hole at the centre of the Milky Way \cite{Eatough}. A magnetic field around a black hole can be present due to, e.g., a pulsar orbiting the black hole, or it can be of the cosmological origin. It has been shown that a black hole located far enough from a magnetar and close its equatorial plane is immersed into a nearly homogeneous magnetic field \cite{Kovar}. A magnetic field makes possible motion of a charged particle with ultrarelativistic velocity. Such a particle emits an electromagnetic radiation whose properties are very similar to the properties of the synchrotron radiation from electrons in accelerators. 

A synchrotron radiation in a strong gravitational field generated by a black hole was studied quite widely. For example, a synchrotron radiation from a charged particle moving in a circular nongeodesic orbit around a weakly magnetized Schwarzschild and Kerr black hole was studied in, e.g., \cite{KhSh} and \cite{Ali,Sok}, respectively. High-frequency electromagnetic and gravitational radiation from a relativistic particle falling into a Schwarzschild and a Kerr black hole was considered and the spectral and angular distributions of the radiation power were calculated in \cite{Dym}. The spectrum of electromagnetic radiation power from radially free falling monopole and point-like dipole into a Schwarzschild black hole was found in \cite{Nov1}. Characteristic features of the electromagnetic spectrum of radiation from a radially free falling dipole were studied subsequently in \cite{Nov2}. An electromagnetic bremsstrahlung spectrum of a dipole falling into a Schwarzschild black hole along a spiral orbit was studied in \cite{Nov3}. An electromagnetic radiation spectrum from charged particles moving slowly in an eccentric bound equatorial orbit around a Schwarzschild black hole immersed in to a weak dipolar magnetic field was derived in \cite{Pap}.

In this paper we shall study a synchrotron radiation form a charged particle moving in the equatorial plane of a Schwarzschild black hole immersed into a uniform magnetic field. The particle is initially moving in a nongeodesic `curly' orbit which due to synchrotron radiation decays to a nongeodesic circular orbit. We shall study this transition and calculate the radiated power and energy loss of the particle.  

This paper is organized as follows: In Sec. II we describe briefly a weakly magnetized Schwarzschild black hole. In Sec. III we review an equatorial motion of a charged particle around a weakly magnetized Schwarzschild black hole. Section IV contains a description of synchrotron radiation in a strong gravitational field. In Sec. V we calculate the energy loss of a charged particle due to its radiation. In Sec. VI we study the synchrotron radiation from a charged particle moving in a bound orbit in the equatorial plane of a weakly magnetized Schwarzschild black hole. Section VII contains a summary of the derived results.

We shall use the conventions adopted in \cite{MTW} and units where $G=c=1$. Occasionally, we shall use the Gaussian system of units.  

\section{Weakly magnetized Schwarzschild black hole}

In this section we briefly review the construction of a weakly magnetized Schwarzschild black hole. An extended review of magnetized black holes is presented in \cite{AG}. To construct a weakly magnetized Schwarzschild black hole of the mass $M$ we assume that there is a weak magnetic field of the characteristic strength $B$, such that the spacetime curvature due to this magnetic field  is much less than the spacetime curvature near the black hole alone, i.e., $GB^{2}/c^{4}\ll c^{4}/G^{2}M^{2}$. This gives
\be\n{0}
B\ll\frac{c^{4}}{G^{3/2}M_{\odot}}\left(\frac{M_{\odot}}{M}\right)\sim10^{19}\left(\frac{M_{\odot}}{M}\right)\text{G}\,,
\ee
where $M_{\odot}$ is the solar mass. In this case, the back reaction of the magnetic field on the spacetime geometry can be neglected. As a result, the spacetime metric is that of the Schwarzschild one,
\ba
ds^2&=&-\left(1-\frac{r_{g}}{r}\right)dt^2+\left(1-\frac{r_{g}}{r}\right)^{-1}dr^2+r^2 d\Omega^2\,,\non\\
d\Omega^2&=&d\theta^2+\sin^2\theta d\phi^2\hhh r_{g}=2M\,.\n{1}
\ea
This spacetime has isometries defined by the Killing vectors $\xi^{\mu}_{(t)}=\delta^{\mu}_{t}$, the generator of time translations, and $\xi^{\mu}_{(\phi)}=\delta^{\mu}_{\phi}$, the generator of azimuthal rotations. Let the magnetic field be static and axisymmetric. Such a magnetic field can be constructed as follows (see, e.g., \cite{Wald,AG}): The Maxwell equation for an electromagnetic 4-potential $A^\mu$ in the Lorenz gauge $A^{\mu}_{\,\,\,;\mu}=0$ reads $A^{\mu;\nu}_{\,\,\,\,\,\,\,;\nu}=R^{\mu}_{\,\,\,\nu}A^{\nu}$. On the other side, a Killing vector obeys the equation $\xi^{\mu;\nu}_{\,\,\,\,\,\,\,\,;\nu}=-R^{\mu}_{\,\,\,\nu}\xi^{\nu}$. Thus, because the metric \eq{1} is Ricci flat, one can take
\be\n{2}
A^{\mu}=\frac{B}{2}\xi^{\mu}_{(\phi)}\,,
\ee
where $B=const$. This 4-vector potential defines a magnetic field which is static, axisymmetric, and homogeneous at the spatial infinity, where it has the constant strength $B>0$ and directed upward, orthogonal to the equatorial plane $\theta=\pi/2$. 

The 4-potential \eq{2} is invariant with respect to the isometries of the Killing vectors, i.e., ${\cal L}_{{\bm \xi}_{(i)}}{\bm A}=0$, $i=(t,\phi)$. The corresponding electromagnetic field tensor $F_{\mu\nu}=A_{\nu,\mu}-A_{\mu,\nu}$ has the following form:
\ba
F_{\mu\nu}&=&2Br\sin\theta\left(\sin\theta\delta^{r}_{[\mu}\delta^{\phi}_{\nu]}+r\cos\theta
\delta^{\theta}_{[\mu}\delta^{\phi}_{\nu]}
\right)\,.\n{3}
\ea

The Schwarzschild metric \eq{1} together with the vector potential \eq{2} define a weakly magnetized Schwarzschild black hole --- a Schwarzschild black hole immersed into a constant uniform magnetic field.  

\section{Equatorial motion of a charged particle}

In this section we review the main results derived in \cite{FS}. A dynamical equation for a charged particle of the mass $m$ and the electric charge $q$ moving in the vicinity of a weakly magnetized Schwarzschild black hole reads
\be
m\frac{D^{2}x^{\mu}}{d\tau^{2}}=qF^{\mu}_{\,\,\nu} u^{\nu}\, .\n{4}
\ee
Here $D/d\tau$ stands for the covariant derivative defined in the metric \eq{1},  $\tau$ is the particle's proper time,
$F_{\mu\nu}$ is given by Eq. \eq{3}, and $u^{\mu}=dx^{\mu}/d\tau$ is the particle 4-velocity, $u^{\mu}u_{\mu}=-1$. For this motion there are two conserved quantities associated with the spacetime Killing vectors: the energy $E>0$ and the generalized azimuthal angular momentum $L_{\phi}\in(-\infty,+\infty)$,
\ba
E&\equiv& -\xi^{\mu}_{(t)}P_{\mu}=m\frac{dt}{d\tau}\left(1-\frac{r_{g}}{r}\right)\,,\n{5a}\\
L_{\phi}&\equiv& \xi^{\mu}_{(\phi)}P_{\mu}=\left(m\frac{d\phi}{d\tau}+\frac{qB}{2}\right)r^2\sin^2\theta\,.\n{5b}
\ea
Here $P_{\mu}=mu_{\mu}+qA_{\mu}$ is the generalized 4-momentum of the particle, where the 4-potential $A^{\mu}$ is given by Eq. \eq{2}. An invariance of the 4-potential with respect to the spacetime isometries [${\cal L}_{{\bm \xi}_{(i)}}{\bm A}=0$, $i=(t,\phi)$] ensures the conservation of $E$ and $L_{\phi}$ along dynamical orbits defined by Eq. \eq{4}. Note that $L_{\phi}$ is defined with respect to the axis $r=0$ where fixed points of the Killing vector $\xi^{\mu}_{(\phi)}$ are located.

One can check that the $\theta$-component of the dynamical equation \eq{4} allows for a solution $\theta=\pi/2$. This is a motion in the equatorial plane of the black hole, orthogonal to the magnetic field. For an equatorial motion  the conserved quantities \eq{5a} and \eq{5b} are sufficient for the complete integrability of the dynamical equation.

In what follows, it is convenient to introduce the dimensionless quantities
\ba
{\cal T}&=&\frac{t}{r_{g}}\hhh\rho=\frac{r}{r_{g}}\hhh \sigma=\frac{\tau}{r_{g}}\,,\non\\
{\cal E}&=&\frac{E}{m}\hhh \ell=\frac{L_{\phi}}{mr_{g}}\hhh b=\frac{qBr_{g}}{2m}\,.\n{6}
\ea
The first integral of the dynamical equation in the equatorial plane takes the following
form: 
\ba
\dot{\cal T}&=&\frac{{\cal E}\rho}{\rho-1}\hhh \dot{\phi}=\frac{\ell}{\rho^2}-b\hhh \dot{\rho}^2={\cal E}^2-U\,,\n{7a}\\
U&=&\left(1-\frac{1}{\rho}\right)\left[1+\frac{(\ell-b\rho^2)^2}{\rho^2}\right]\,,\n{7b}
\ea
where $U$ is the effective potential. Here and in what follows, the over-dot stands for the derivative with respect to $\sigma$. The effective potential is positive in the domain of interest $\rho\in(1,+\infty)$. It vanishes at $\rho\to1$ and diverges at $\rho\to+\infty$. Depending on values of $\ell$ and $b$ the potential is either monotonically increasing or has two extrema, a local maximum and a local minimum. 

Equations \eq{7a} are invariant with respect to the discrete transformations
\be\n{8}
b\to-b\hhh \ell\to-\ell\hhh\phi\to-\phi\,.
\ee
According to the direction of the magnetic field ($B>0$), for a positive electric charge $q$ we have $b>0$. Without loss of the generality, we can take $q>0$. Equations for a particle with a negative charge are derived from those for a particle with a positive charge by the transformation \eq{8}.

The parameter $\ell$ can be positive or negative. For $\ell>0$ the Lorentz force acting on a charged particle is repulsive, i.e., it is directed outward from the black hole. For $\ell<0$ the Lorentz force is attractive, i.e., it is directed toward the black hole. In what follows, we shall consider the case when $\ell>0$.

Here we shall consider bound orbits.\footnote{Bound orbits of small harmonic radial and latitudinal oscillations around a stable circular orbit were studied in \cite{Kolos}.} A study of such orbits may help to analyze motion of particles in a black hole accretion disk. Bound orbits exist if the effective potential has two extrema: ${\cal E}^{2}_{\text{max}}=U(\rho_{\text{max}})$ and ${\cal E}^{2}_{\text{min}}=U(\rho_{\text{min}})$. For a bound orbit the radial coordinate $\rho$ oscillates between the minimal $\rho_{1}$ and the maximal $\rho_{2}$ values, such that $\rho_{\text{max}}\leq\rho_{1}\leq\rho_{\text{min}}\leq\rho_{2}$ and for a particular bound orbit the particle's energy ${\cal E}$ obeys the relation ${\cal E}_{\text{min}}\leq{\cal E}\leq{\cal E}_{\text{max}}$. Equation \eq{7a} for the azimuthal coordinate $\phi$ implies that there is a critical value $\rho_{\text{cr}}=\sqrt{\ell/b}>\rho_{\text{min}}$ for which $\dot{\phi}=0$. For $\rho<\rho_{\text{cr}}$ one has $\dot{\phi}>0$ and for $\rho>\rho_{\text{cr}}$ one has $\dot{\phi}<0$.  As a result, there are two different types of bound orbits separated by the critical one. The first type corresponds to $\rho_{2}<\rho_{\text{cr}}$ and looks similar to a contracted trochoid. The second type corresponds to $\rho_{2}>\rho_{\text{cr}}$ and has loops. It looks similar to a prolate trochoid. The critical type is defined by $\rho_{2}=\rho_{\text{cr}}$ and has cusps. It looks similar to a cycloid (see Fig.~\ref{F3} below). 

\section{Synchrotron radiation}

For an ultrarelativistic  particle of a charge $q$ moving in Minkowski spacetime the radiated four-momentum is given by
\be\n{12}
 d{\cal P}^{\hal}=\frac{2q^{2}}{3}\frac{d^{2}x^{\hmu}}{d\tau^{2}}\frac{d^{2}x_{\hmu}}{d\tau^{2}}u^{\hal}d\tau\,,
 \ee
where $x^{\hal}$ are natural (Minkowski) coordinates (see, e.g. \cite{Lan}). Because of nonlocal nature of the process of formation of electromagnetic radiation, this expression generally is not applicable for a charged particle moving in a curved spacetime. This is the case when the particle moves along a spacetime geodesic and the characteristic length $\delta l$ of the particle's orbit where the dominant part of the radiation is formed is of the order of the characteristic scale of the gravitational field $L\sim r_{g}$. However, if dynamics of the particle is mostly governed by forces of nongravitational, e.g. electromagnetic, nature such that the particle's world line is not a geodesic, then the process of formation of electromagnetic radiation can approximately be described by local quantities (see, e.g., \cite{KhSh,STAG,Gal}).       
 
To see why such a description is possible, let us consider a locally geodesic frame of reference localized somewhere in the vicinity of the midpoint of $\delta l$. Such a frame of reference can be defined by the normal Riemann coordinates $y^{\hal}$ (see, e.g., \cite{MTW}). In this frame we have
\ba
g_{\hmu\hnu}&=&\eta_{\hmu\hnu}+{\cal O}(y^{2})\,,\n{13a}\\
\Gamma^{\hal}_{\,\,\,\hmu\hnu}&=&-\frac{1}{3}(R^{\hal}_{\,\,\,\hmu\hnu\hgam}+R^{\hal}_{\,\,\,\hnu\hmu\hgam})y^{\hgam}+{\cal O}(y^{2})\,,\n{13b}
\ea
where $\eta_{\hmu\hnu}$ is an orthonormal metric and $y$ is the proper size of the spacetime domain in the vicinity of the particle location covered by the normal Riemann coordinates. Let us now estimate the characteristic wavelength of a synchrotron radiation, $\lambda_{*}$. The angular distribution of the synchrotron radiation is localized in a narrow cone of the opening angle $\delta\varphi\approx\hgam^{-1}\ll1$ around the direction of the particle velocity (see, e.g., \cite{Lan,Ternov}). This ``projector effect'' is characteristic for a synchrotron radiation. Let $r_{c}$ be the local curvature radius of the particle orbit. Then, $\delta l\approx r_{c}\hgam^{-1}$. We shall now find the characteristic wavelength of the radiation. One has $\lambda_{*}=\hT$, where $\hT$ is the period of the radiated wave measured by a local observer. On the other side, $\delta l=\delta\htt'=(d\htt'/d\htt)\hT$, where $\htt'$ is the time measured in the particle's frame and $\htt$ is the time measured by the observer. To calculate $d\htt'/d\htt$ we shall use the relation $\htt=\htt'+\hR(\htt')$, where  $\hR(\htt')$ is the distance measured from the particle to the observer along the light trajectory. Using the obvious expression $d\hR/d\htt'=-\hat{\BM n}\cdot\hat{\BM v}=\hv\cos\delta\varphi$, where $\hat{\BM n}$ is the unit vector in the direction of the radiation, we derive
\be\n{14}
\frac{d\htt}{d\htt'}\approx1-\hv+\frac{\hv}{2}\delta\varphi^{2}\approx\hgam^{-2}\,.
\ee
As a result, we find
\be\n{15}
\lambda_{*}\approx\delta l\hgam^{-2}\approx r_{c}\hgam^{-3}\,.
\ee
This expression implies that $\lambda_{*}\ll r_{c}$. It means that the wave zone of the radiation begins at the distances shorter than $r_{c}$. 

The wave equation for the radiation 4-vector potential $A^{\alpha}_{\text{rad}}$ reads
\be\n{15a}
-A^{\alpha;\beta}_{\text{rad}\,\,\,;\beta}+R^{\alpha}_{\,\,\,\beta}A^{\beta}_\text{rad}=4\pi J^{\alpha}\hhh A^{\alpha}_{\text{rad}\,;\alpha}=0\,.
\ee
In a locally geodesic frame, in a vacuum spacetime, with the use of the expressions \eq{13a}--\eq{13b} it takes the following linear in $y^{\hal}$ form:
\ba\n{16a}
&&-\square A^{\hal}_{\text{rad}}+\frac{2}{3}\left(R^{\hal\hmu}_{\,\,\,\,\,\,\,\hnu\hgam}+R^{\hal\,\,\,\hmu}_{\,\,\,\hnu\,\,\,\,\hgam}\right)y^{\hgam}A^{\hnu}_{\text{rad}\,,\hmu}\approx4\pi J^{\hal}\,,\non\\
&&A^{\hal}_{\text{rad}\,;\hal}\approx0\,,
\ea
where $\square$ is the wave operator defined in the orthonormal metric $\eta_{\hmu\hnu}$. For a synchrotron radiation the terms containing the Riemann tensor components are of the relative order of $\lambda_{*} y/L^{2}$. For $y\lesssim L$ the relative order of the terms proportional to the Riemann tensor due to large Lorentz gamma factor is small even for $r_{c}\sim L$. Thus, such terms can be neglected and the wave equation can be approximated by that in a flat spacetime.   

As a result, the synchrotron radiation from an electrically charged particle moving in a non-geodesic world line in the vicinity of a weakly magnetized Schwarzschild black hole can be described in a locally geodesic frame of reference where the spacetime is practically flat and the expression \eq{12} is applicable.  Note, however, that this expression correctly covers the entire range of frequencies in flat spacetime only. In a curved spacetime the low frequency part depends on the gravitational field which is not included into consideration. As a result, the low frequency part is neglected when one uses the expression \eq{12} in a curved spacetime.  

In the global Schwarzschild coordinates the radiated four-momentum of a charged particle moving around a weakly magnetized Schwarzschild black hole, as measured by a static observer located in the asymptotically flat region, reads
\be\n{16}
d{\cal P}^{\alpha}=\frac{2q^{2}}{3}\frac{D^{2}x^{\mu}}{d\tau^{2}}\frac{D^{2}x_{\mu}}{d\tau^{2}}dx^{\alpha}\,.
\ee 

For the dynamical motion \eq{4} this expression can be written as
\be\n{17}
d{\cal P}^{\alpha}=\frac{2q^{4}}{3m^{2}}F^{\mu}_{\,\,\,\nu} u^{\nu}F_{\mu\lambda} u^{\lambda}dx^{\alpha}\,.
\ee
For an equatorial motion $(\theta=\pi/2)$ in the ultrarelativistic approximation $\gamma=dt/d\tau\gg1$, the radiated four-momentum \eq{17} reads
\be\n{18}
d{\cal P}^{\alpha}=\frac{2q^{4}B^{2}}{3m^{2}}\gamma^{2}\left(1-\frac{r_{g}}{r}\right)^{2}dx^{\alpha}\,.
\ee

One can also calculate the spectral distribution function for the synchrotron radiation. Introducing the radiation intensity 
\be\n{18aa}
I=-dE/dt\,,
\ee
where $E$ is the charged particle's energy, one can derive the following expression (for details see \cite{STAG,Gal}): 
\be\n{18a}
\frac{dI}{dy}=\frac{3\sqrt{3}}{4\pi}\frac{q^{4}B^{2}}{m^{2}}\gamma^{2}\left(1-\frac{r_{g}}{r}\right)^{3}F(y)\,,
\ee        
where 
\be\n{18b}
y=\frac{\Omega}{\Omega_{c}}\hhh \Omega_{c}=\frac{3qB}{2m}\gamma^{2}\left(1-\frac{r_{g}}{r}\right)^{2}\,,
\ee
and
\be\n{18c}
F(y)=y\int_{y}^{\infty}K_{5/3}(x)dx\hhh\int_{0}^{\infty}F(y)dy=\frac{8\pi}{9\sqrt{3}}\,.
\ee
Here $K_{5/3}(x)$ is the MacDonald function and $\Omega$ is the radiation frequency. The function $F(y)$ is localized hear its maximum at $y\approx0.29$. Thus, the main part of the synchrotron radiation is concentrated around the characteristic frequency
\be\n{18d}
\Omega_{*}\approx0.29\Omega_{c}\,.
\ee
The total radiation intensity $I$ is derived by integrating the expression \eq{18a} over $y\in[0,\infty)$,
\be\n{18e}
I=\frac{2}{3}\frac{q^{4}B^{2}}{m^{2}}\gamma^{2}\left(1-\frac{r_{g}}{r}\right)^{3}\,.
\ee
   
\section{Energy loss}

The expression \eq{18e} allows us to calculate the rate of the energy loss measured with respect to the particle's proper time. Using the expressions \eq{6} and \eq{7a} we derive the rate of the energy loss, 
\be\n{19}
\dot{\cal E}=-\kappa{\cal E}^{3}\hhh \kappa=\frac{8q^{2}b^{2}}{3mr_{g}}\,.
\ee 
Integrating we derive
\be\n{20}
{\cal E}(\sigma)=\frac{{\cal E}_{i}}{\sqrt{1+2\kappa{\cal E}_{i}^{2}\sigma}}\,,
\ee
where ${\cal E}_{i}$ is the initial energy. This expression illustrates how the particle's energy decreases with its proper time due to the synchrotron radiation. The rate of the energy decrease is controlled by the parameter $\kappa$.

In order to estimate the value of $\kappa$, we shall use the Gaussian system of units. We have
\be\n{21}
\kappa\approx191\left(\frac{q}{e}\right)^{4}\left(\frac{m}{m_{e}}\right)^{-3}\left(\frac{B}{10^{8}\text{G}}\right)^{2}\left(\frac{M}{M_{\odot}}\right)\,,
\ee
where $e$ is the elementary electric charge and $m_{e}$ is the electron mass. According to Piotrovich, Silant'ev, Gnedin, and Natsvlishvili \cite{MagneticField}, the characteristic scales of the magnetic field $B$ are of the order of $10^8$G near the horizon of a stellar mass black hole, $M\sim 10 M_{\odot}$, and of the order of $10^4$G  near the horizon of a supermassive black hole, $M\sim 10^9 M_{\odot}$. As a result, we derive $\kappa\approx1.910\times10^{3}$ for an electron orbiting a stellar mass or a supermassive black hole, and $\kappa\approx3.085\times10^{-7}$ for a proton orbiting a stellar mass or a supermassive black hole. 

Thus, ultrarelativistic electrons orbiting astrophysical black holes loose their energy due to their synchrotron radiation relatively fast. On the other side, for protons or heavy ions orbiting astrophysical black holes $\kappa\ll1$. This implies that protons and heavy ions loose their energy relatively slow.

\section{Synchrotron radiation from bound orbits}

\begin{figure}
\includegraphics[width=6cm]{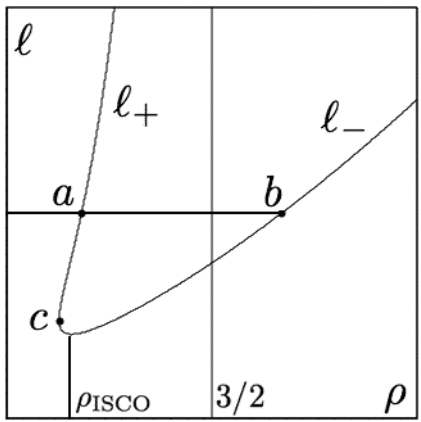}\\
\caption{The branches $\ell_{\pm}>0$. The branches $\ell_{+}$ and $\ell_{-}$ merge at the point $c$. For sufficiently large values of $\ell=\ell_{+}(\rho_{\text{max}})=\ell_{-}(\rho_{\text{min}})$ the points $a$ and $b$ correspond to the maximum and minimum of the effective potential, respectively. The maximum and minimum of the effective potential merge at $\rho=\rho_{\text{ISCO}}$ where $\ell_{-}$ has the minimal value.} \label{F1}
\end{figure}

The shape of the effective potential \eq{7b} is controlled by values of the parameters $b$ and $\ell$. For a given weakly magnetized Schwarzschild black hole and a charged particle the parameter $b$ is fixed [see Eq. \eq{6}]. In order to have bound orbits (which are possible if the effective potential has two extrema) the dynamical parameter $\ell$ should take certain values related to the fixed value of $b$. Let us estimate these values. In the Gaussian system of units the parameter $b$ has the following value:
\be\n{29}
b\approx8.663\times10^{9}\left(\frac{q}{e}\right)\left(\frac{m}{m_{e}}\right)^{-1}\left(\frac{B}{10^{8}\text{G}}\right)\left(\frac{M}{M_{\odot}}\right)\,.
\ee 
According to \cite{MagneticField}, we have $b\approx4.718\times10^{7}$ for a proton orbiting a stellar mass black hole and $b\approx4.718\times10^{11}$ for a proton orbiting a supermassive black hole. For an electron the value of $b$ is $m_{p}/m_{e}\approx1836$ times larger. In order to find the corresponding value of $\ell$ we use the equation for extrema of the effective potential, $U_{,\rho}=0$. This equation is equivalent to the equation
\be\n{29a}
(3-2\rho)\ell^{2}-2b\rho^{2}\ell+\rho^{2}(2b^{2}\rho^{3}-b^{2}\rho^{2}+1)=0\,,
\ee
which is a quadratic polynomial in $\ell$. Solving this equation for $\ell$ we derive
\be\n{29b}
\ell_{\pm}=\frac{b\rho^{2}\pm\rho\sqrt{4b^{2}\rho^{2}(\rho-1)^{2}-3+2\rho}}{3-2\rho}\,.
\ee
The solution $\ell_{-}$ is positive in the domain $\rho\in[\rho_{c},+\infty)$, where $\rho_{c}\in(1,3/2)$. The solution $\ell_{+}$ has two separate branches. The branch corresponding to $\ell_{+}>0$ is defined in the domain $\rho\in[\rho_{c},3/2)$. In what follows, according to our choice ($\ell>0$), we shall consider this branch of $\ell_{+}$. This branch and the branch $\ell_{-}$ merge at $\rho=\rho_{c}$. For large values of $b$, $\ell=\ell_{\pm}$ is large as well. In this case, the branch $\ell_{-}$ corresponds to the minimum of $U$, $\ell=\ell_{-}(\rho_{\text{min}})$ and the branch $\ell_{+}$ corresponds to the maximum of $U$, $\ell=\ell_{+}(\rho_{\text{max}})$. The branches $\ell_{\pm}$ are schematically illustrated in Fig.~\ref{F1}. 

Accordingly, for large values of $b$ one can derive the following expansions of the expression \eq{29b}:
\ba
\ell_{+}&=&\frac{b\rho_{\text{max}}^{2}(2\rho_{\text{max}}-1)}{(3-2\rho_{\text{max}})}+{\cal O}(b^{-1})\,,\n{30a}\\
\ell_{-}&=&b\rho_{\text{min}}^{2}+\frac{1}{4b(\rho_{\text{min}}-1)}+{\cal O}(b^{-3})\,.\n{30b}
\ea

In the frame of reference where the charged particle is at rest, the dipole radiation is symmetric and, therefore, the radiated momentum is zero. However, in another frame of reference, e.g., in a locally geodesic frame, the dipole radiation is mostly concentrated in a narrow cone in the direction of the particle velocity. Thus, the synchrotron radiation carries away the azimuthal angular momentum of a charged particle. The expression \eq{18} allows us to calculate the rate of the angular momentum loss measured with respect to the particle's proper time, $dL_{\phi}/d\tau=-g_{\phi\phi}d{\cal P}^{\phi}/d\tau$. Using the expressions \eq{6} and \eq{7a} we derive
\be\n{31}
\dot\ell=-\kappa{\cal E}^{2}\rho^{2}\dot\phi=-\kappa{\cal E}^{2}(\ell-b\rho^{2})\,.
\ee
Having this expression we can compare the relative rate of the energy and the angular momentum loss. For a charged particle moving in the vicinity of the black hole we have $\rho\sim1$ and the expression \eq{30b} gives $\ell\sim b$. Using the expressions \eq{19} and \eq{31} we derive
\be\n{32}
\frac{|\dot\ell|}{\ell}\sim\frac{|\dot{\cal E}|}{{\cal E}}b^{-2}\,.
\ee
Thus, for relatively short time intervals one can neglect the angular momentum loss. In what follows, we shall consider $\ell\approx const$.

\begin{figure}
\includegraphics[width=6cm]{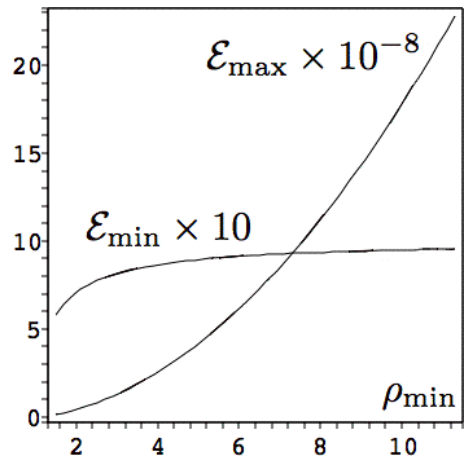}\\
\caption{The minimal and the maximal values of a proton's energy which is moving around a stellar mass black hole in a bound orbit oscillating about the fiducial stable circular orbit of the radius $\rho_{\text{min}}$.} \label{F2}
\end{figure}

\begin{figure}
\includegraphics[width=6cm]{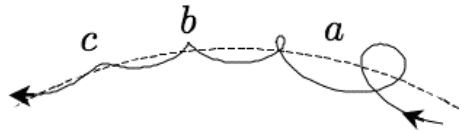}\\
\caption{Schematic illustration of the decay of radial oscillations: a transition from the ``prolate trochoid'' orbit (section $a$) to ``cycloid'' (section $b$) and to ``contracted trochoid'' (section $c$). The dashed line represents the fiducial stable circular orbit of $\rho=\rho_{\text{min}}$.} \label{F3}
\end{figure}

Using Eqs. \eq{7b}, \eq{30a}, and \eq{30b} one can find the corresponding expansions of the extrema,\footnote{Note that $\ell_{+}$ and ${\cal E}_{\text{max}}^{2}$ diverge at $\rho_{\text{max}}=3/2$. This value of $\rho_{\text{max}}$ corresponds to an unstable circular orbit of the radius $r=3M$, which is a null geodesic of the Schwarzschild spacetime.}
\ba
&&\hspace{-0.5cm}{\cal E}_{\text{max}}^{2}=\frac{16b^{2}\rho_{\text{max}}(\rho_{\text{max}}-1)^{3}}{(3-2\rho_{\text{max}})^{2}}+{\cal O}(1)\,,\n{33a}\\
&&\hspace{-0.5cm}{\cal E}_{\text{min}}^{2}=1-\frac{1}{\rho_{\text{min}}}+\frac{1}{16b^{2}\rho_{\text{min}}^{3}(\rho_{\text{min}}-1)}+{\cal O}(b^{-4})\,.\n{33b}
\ea
The maximal and the minimal values of a proton's energy as functions of $\rho_{\text{min}}$ are presented in Fig.~\ref{F2}. One can see that ${\cal E}_{\text{max}}$ is of several orders higher than ${\cal E}_{\text{min}}$. For a given value of $\rho_{\text{min}}$ which defines a stable circular orbit about which a charged particle moving in a bound orbit oscillates, one can find the difference ${\cal E}_{\text{max}}-{\cal E}_{\text{min}}$ which is a measure of the energy carried out by synchrotron radiation. As one can see from Fig.~\ref{F2}, this energy is of the order of ${\cal E}_{\text{max}}$.

A decrease of the energy of a charged particle moving in a bound orbit results in a transition from the ``prolate trochoid'' orbit to ``cycloid'' and to ``contracted trochoid'' followed by decay of the radial oscillations, so that the particle orbit becomes nearly circular. This transition is schematically illustrated in Fig.~\ref{F3}. Note that a real prolate trochoid orbit has a very large number of densely packed loops (see \cite{FS}).  

Let us now calculate the dimensionless proper time $\Delta\sigma$ required to lower the particle energy from ${\cal E}_{i}={\cal E}_{\text{max}}$ to ${\cal E}_{f}$. Using the expression \eq{20}, we derive
\be\n{33}
\Delta\sigma=\frac{{\cal E}_{i}^{2}-{\cal E}_{f}^{2}}{2\kappa{\cal E}_{i}^{2}{\cal E}_{f}^{2}}\,.
\ee
One can estimate an upper bound of $\Delta\sigma$. Taking ${\cal E}_{i}={\cal E}_{\text{max}}$ and ${\cal E}_{f}={\cal E}_{\text{min}}$ and using Eq. \eq{33b} we derive
\be\n{34}
\Delta\sigma\lesssim\Delta\sigma_{\text{max}}\approx\frac{\rho_{\text{min}}}{2\kappa(\rho_{\text{min}}-1)}\,.
\ee
The corresponding to $\Delta\sigma$ portion of the fiducial stable circular orbit $\Delta{\cal C}$ about which the particle makes the radial oscillations can be calculated with the use of its drift angular velocity [Cf. Eqs. \eq{7a} and \eq{30b}],
\be\n{35}
\langle\dot\phi\rangle\approx\frac{\ell}{\rho_{\text{min}}^{2}}-b\approx\frac{1}{4b\rho_{\text{min}}^{2}(\rho_{\text{min}}-1)}\,.
\ee 
We have
\be\n{36}
\Delta{\cal C}=\Delta\sigma\frac{\langle\dot\phi\rangle}{2\pi}\approx\frac{\Delta\sigma}{8\pi b\rho_{\text{min}}^{2}(\rho_{\text{min}}-1)}\,.
\ee
The maximal portion of the fiducial circular orbit corresponds to $\Delta\sigma_{\text{max}}$,
\be\n{37}
\Delta{\cal C}_{\text{max}}\approx\frac{1}{16\pi\kappa b\rho_{\text{min}}(\rho_{\text{min}}-1)^{2}}\,.
\ee

\begin{figure}
\includegraphics[width=6cm]{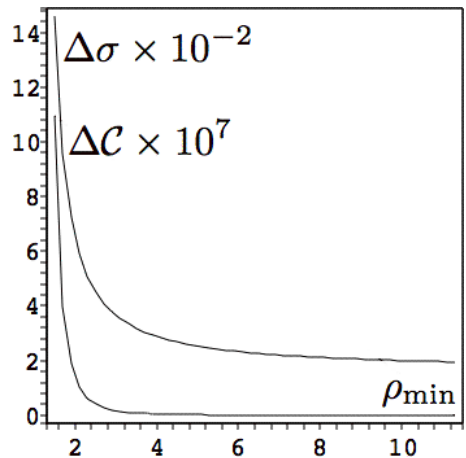}\\
\caption{The dimensionless proper time $\Delta\sigma$ and the portion of the fiducial circular orbit $\Delta{\cal C}$ for a proton orbiting a stellar mass black hole. The initial and final energy values of the proton are ${\cal E}_{i}={\cal E}_{\text{max}}$ and ${\cal E}_{f}=\gamma(\rho_{\text{min}}-1)/\rho_{\text{min}}$ with $\gamma=100$.} \label{F4}
\end{figure}

The quantities defined above are presented in Fig.~\ref{F4} as functions of $\rho_{\text{min}}$. One can see that the dimensionless proper time $\Delta\sigma$ as well as the portion of the fiducial circular orbit $\Delta{\cal C}$ rapidly decrease for bound orbits corresponding to larger values of $\rho_{\text{min}}$. This implies that an intensity of the synchrotron radiation is considerably greater from particles moving in bound orbits of larger radii. Note that because the value of $\kappa$ for an electron is of many orders larger than that of a proton or a heavy ion, the corresponding values of $\Delta\sigma$ and $\Delta{\cal C}$ are of many orders smaller. Thus, bound orbits of electrons decay much faster than bound orbits of protons and heavy ions. 

Let us finally calculate an evolution of the characteristic frequency of the synchrotron radiation. Using the expressions \eq{6}, \eq{7a}, \eq{18b}, and \eq{18d} we derive the dimensionless characteristic frequency,
\be\n{38}
\omega_{*}=\Omega_{*}r_{g}\approx0.87b{\cal E}^{2}\,.
\ee
Then, the expression \eq{20} gives,
\be\n{39}
\omega_{*}\approx\frac{0.87b{\cal E}_{i}^{2}}{1+2\kappa{\cal E}_{i}^{2}\sigma}\,.
\ee
The upper bound on the characteristic frequency is [Cf. Eq. \eq{33a}]
\be\n{40}
\omega_{*\text{max}}\approx14b^{3}\frac{\rho_{\text{max}}(\rho_{\text{max}}-1)^{3}}{(3-2\rho_{\text{max}})^{2}}\,.
\ee
According to Fig.~\ref{F2}, the expression \eq{39} implies that the characteristic frequency of the synchrotron radiation is larger for ``prolate trochoid'' type bound orbits oscillating about the fiducial stable circular orbit of the radius $\rho_{\text{min}}$ than that for nearly circular orbits of the same $\rho_{\text{min}}$ value. Moreover, the frequency is larger for the orbits of larger value of $\rho_{\text{min}}$. The rate of the frequency decrease is considerably larger for electrons than that for protons and heavy ions.

\section{Summary}

Ultrarelativistic charged particles orbiting a magnetized Schwarzschild black hole loose their energy due to synchrotron radiation. For electrons the rate of the energy loss measured with respect to the particle's proper time is of many orders larger than that for protons and heavy ions. Moreover, an intensity of the synchrotron radiation is considerably greater from particles moving in bound orbits of larger radii. Due to the energy loss, the radial oscillations about a stable circular fiducial orbit decay. Such a decay results in a transition from a ``prolate trochoid'' bound orbit, which has loops, to ``cycloid'' bound orbit, and finally to ``contracted trochoid'' bound orbit. The transition time, as measured with respect to a particle's proper time, is of many orders shorter for electrons than that for protons and heavy ions. The characteristic frequency of the synchrotron radiation is larger for a charged particle moving in a ``prolate trochoid'' type orbit than that for the charged particle moving in a ``cycloid'' and ``contracted trochoid'' type orbit. This frequency is of approximately 1836 times larger for electrons than that for protons moving along the same type of bound orbit and the rate of the frequency decrease is considerably larger for electrons than that for protons and heavy ions. The derived results may be useful in the model of a hypothetical thin accretion disk around an astrophysical black hole which is composed of charged test particles.

In the analysis presented in this paper the radiation effects were considered on ``a background of given dynamic'' which defines fixed-energy bound orbits. It would be interesting to consider a dynamical analysis where the radiation reaction forces are included into the particle's dynamic.   

\acknowledgments

The author is grateful to the Natural Sciences and Engineering Research Council of Canada for its support.

\end{document}